\documentclass[12pt]{article}
\usepackage{epsfig}
\def\om{\vec{\omega}}

\def\d{{\bf d}}
\def\rn{r_{\nu}}

\def\k{{\bf k}}
\def\kp{\k_{\perp}}
\def\kn{k_{\parallel}}
\def\n{{\bf n}}
\def\u{{\bf U}}

\def\th{{\vec{ \theta}}}

\def\HI{{\rm HI}}

\def\d{{\bf d}}
\begin{document}
\title{HI Fluctuations at Large Redshifts: II - the  Signal Expected 
  for GMRT. }
\author{Somnath Bharadwaj \footnote{Corresponding author} 
\\ Department of Physics and Meteorology \\ \& \\ 
  Center for Theoretical Studies,\\
I.I.T. Kharagpur, 721 302, India \\
email:\,  somnath@phy.iitkgp.ernet.in \\ \\ 
 Sanjay K. Pandey\\
 Deptt. of Mathematics \\
L.B.S.College, Gonda 271001 , India}
\date{}
\maketitle
\begin{center}
Abstract
\end{center}
For the GMRT, we calculate the expected signal from redshifted HI
emission at two frequency bands centered at 610 and 325 MHz. 
The study focuses on  the visibility-visibility cross-correlations, proposed
earlier as the optimal  statistical  estimator for detecting and
analyzing this signal. These correlations directly probe the power
spectrum of density fluctuations at the redshift where the radiation
originated, and thereby provide a method for studying the large scale
structures at large redshifts. We present detailed estimates of the
correlations expected between the visibilities measured at different
baselines and frequencies. Analytic fitting formulas representing the
salient features of the expected signal are also provided. These will
be useful in planning  observations and deciding  an optimal
strategy for detecting this signal. 
\section{Introduction}
In two earlier papers (Bharadwaj, Nath \& Sethi, 2001; hereafter
referred to as Paper I and Bharadwaj  \& Sethi, 2001; hereafter
referred to as Paper II)  we have studied the possibility of
detecting  redshifted 21 cm  emission from neutral hydrogen (HI) at
high redshift sand  using these observations to probe large scale
structures. 
Our approach has been based on the fact that the HI density in the
redshift range $1 \le z \le 3.5$ is known from observations   
of Lyman-$\alpha$ absorption lines  seen in quasar spectra. 
These observations currently indicate $\Omega_{gas}(z)$, the comoving
density of neutral gas expressed as a fraction of the present critical 
density, to be 
nearly constant at a value   $\Omega_{gas}(z) \sim 10^{-3}$ for 
 $z \ge 1$ (Peroux {\it et al.} 2001).
The bulk of the neutral gas is in clouds which have  HI column   
densities greater than $2 \times 10^{20} {\rm atoms/cm^{2}}$ (Peroux
{\it et al.} 2001, Storrie-Lombardi, McMahon, Irwin 1996, Lanzetta, Wolfe,
\& Turnshek 1995). These  high column density clouds are responsible for
the damped Lyman-$\alpha$ absorption lines  observed  along lines  of
sight to quasars. The flux of HI emission from individual clouds
($ < 10 \mu {\rm Jy}$) is too weak to be detected by existing
radio telescopes unless the image of the cloud is   significantly
magnified  by an intervening cluster gravitational lens (Saini,
Bharadwaj and Sethi, 2001). Although we may not be able to detect
individual clouds, the redshifted HI emission from the distribution of  
clouds will appear as  a background radiation in  low frequency radio 
observations. In the two earlier papers, and in the present paper we
investigate issues related to calculating  the expected signal and
detecting it. 

The hope of detecting this signal lies in the fact that the
distribution of the HI clouds  is expected to be 
clustered. The clustering pattern of the HI  clouds will be reflected
in the angular and spectral distribution of the the redshifted 21 cm
emission. It is the aim of this paper, and the two previous papers, to
characterize the expected properties of this signal and to propose a
optimum statistical estimator to extract this signal from the
various foregrounds and the system noise which will also be present in
any low frequency observation.   In paper II we proposed the
cross-correlations between the visibilities measured at different
baselines and frequencies as a statistical estimator 
to optimally detect the HI signal. This has the advantage that the
system noise contribution to  visibilities measured at
different baselines and frequencies are uncorrelated. Further, in
Paper II we showed that the visibility-visibility cross-correlations
directly probes the power spectrum of density fluctuations at the epoch
where the HI emission originated. In this paper we present results of
a detailed investigation of the signal expected for this statistical
estimator for observations with the  GMRT. 

We next present a brief outline of this paper.  In section 2 we
review the formalism used to calculate the visibility-visibility
cross-correlations. In section 3. we present the values of a set of
useful parameters  to do with the GMRT (Swarup {\it et al.}
1990), the cosmological model chosen for these
investigations and the HI distribution. In Section
4. we present the results of our investigations, and in Section 5 we
discuss the results and present conclusions.

\section{Formalism.}
We consider a situation where  an interferometric array of $N$
antennas is  used to carry out low frequency radio
observations. The observations are  carried out at  $NC$
frequency channels $\{ \nu_1,\nu_2,\nu_3, ...,\nu_{NC}\}$ covering a
frequency band   $B$ centered at the frequency $\nu_c$.  For the
purpose of this paper we  
assume that the  antennas are distributed on a plane, and  the
position of each antenna is denoted by a two dimensional vector
$\d_i$.  The antennas all point in the same direction along the unit
vector $\n$ which we take to be 
vertically up wards. The beam pattern $A(\th)$  quantifies how the
individual antenna, pointing upwards,  responds to signals from
different directions in the sky. This is assumed to be a Gaussian
$A(\th)=e^{-\theta^2/\theta_0^2}$ with $\theta_0 \ll 1$ {\it i.e.} the
beam width of  the antennas is small, and the part of the sky which
contributes to the signal can be  well approximated by a plane. 

The quantity  measured in interferometric  observations is the
visibility 
$V(\u,\nu)$ which is recorded for every independent pair of antennas
(baseline) at every frequency channel in the band. For any pair of
antennas, the visibility depends on the vector $\d =\d_i - \d_j$
joining the position of  the two antennas. It is  convenient  to
express the visibility as a function of the variable $\u$ which is
$\d$    expressed  in units of the  wavelength {\it i.e.}
$\u=\d/\lambda$. Typically, the  visibility is 
recorded for $\u$ in a range $U_{\rm min} \le \mid \u \mid 
\le U_{\rm max}$. The visibility corresponding to $\u=0$  is the flux
measured by a single antenna, and this is usually not recorded in an
interferometric array. 

In paper II we calculated the HI contribution to the
visibility $V(\u,\nu)$. This can be expressed as 
\begin{eqnarray}
 V(\u,\nu) =    \bar{I}_{\nu}  \int \frac{d^3 k}{(2 \pi)^3}  
\Delta^{s}_{\HI}(\k,\nu) 
e^{-i  \kn \rn}  \,  a_{\nu}(\u-\frac{\kp \rn}{2 \pi}) \,.
\label{eq:a1}
\end{eqnarray} 

The frequency $\nu$ appears in many of the terms on the right hand
side of equation (\ref{eq:a1}) to indicate that these are to be
evaluated at the redshift $z=\nu_e/\nu-1$ where the radiation
originated. 
The first term $ \bar{I}_{\nu}$ is the specific intensity of the
redshifted HI emission expected if the HI were uniformly
distributed. This is given to be   (from eq. (3) of Paper II) 
\begin{equation}
\bar{I}_{\nu}=\frac{2.7 \,  {\rm Jy}}{\rm degree^2} \,\,  h \,
 \Omega_{\rm gas}(z) \, \, \left[\frac{H_0}{H(z)} \right] 
\label{eq:a2}
\end{equation}
where $H(z)$ is the Hubble parameter, and  $H_0$ is its present value. 

It should be noted that eq. (14) of Paper II is incorrect. Equation~3
of paper II is correct.  The error came about when expressing 
$\bar{n}_{\rm HI}(z)$, the comoving number  density of HI atoms   in
terms of $\Omega_{\rm HI}$. As a result of this and an error  in the
computer programs that were used, the predictions  for $610 \, {\rm MHz}$
in Paper II are also incorrect.

The  HI emission originates from the gas  located at a  distance  $r_{\nu}$
in the direction $\n$. The  term $\Delta^s_{\HI}(\k,\nu)$ in equation
(\ref{eq:a1}) is the 
Fourier transform of the density contrast  of the HI distribution in
redshift space. Any Fourier mode $\k$ can be decomposed into two
parts, a component $\kn$ parallel to $\n$ and a component 
$\kp$ perpendicular $\n$. Assuming that the 
HI density fluctuations  evolve according to linear theory, and that
these may be related  to the fluctuations in  the underlying dark
matter distribution through a linear bias  parameter $b(z)$,  we have  
\begin{equation}
\Delta^s_{\HI}(\k,\nu)=b(z) D(z) \left[ 1 +
  \beta(z) \frac{\kn^2}{k^2} \right] \Delta(\k) 
\label{eq:a3}
\end{equation}
where $D(z)$ is the growing mode of density perturbations (Peebles, 1980),
$\beta(z)=f(\Omega_m,\Omega{\Lambda})/b(z)$ (Lahav {\it et al.}
1991)   is the linear redshift distortion parameter, and $\Delta(\k)$ 
is the Fourier transform of the density fluctuations of the dark
matter distribution at the present epoch.  

We also have 
\begin{equation}
\langle \Delta(\k) \Delta^{``}(\k^{'}) \rangle =
(2 \pi)^3 \delta^3(\k-\k^{'}) P(k)
\label{eq:a4}
\end{equation}
where $\langle \rangle$ denotes ensemble average, and  $P(k)$ is the
power-spectrum of  dark matter fluctuations at the present epoch. 

The quantity $a_{\nu}(\u)$ in equation
 (\ref{eq:a1}) is the Fourier  transform of   $A(\theta)$, the beam
 pattern.  For a Gaussian 
 beam, the Fourier transform also is a  Gaussian and we have   
\begin{equation}
a_{\nu}(\u-\frac{\kp \rn}{2 \pi})=
\pi \theta_{0\nu}^2 \exp{\left[-\pi^2 \theta_{0 \nu}^2
 \left(\u-\frac{\kp   \rn}{2 \pi} \right)^2
    \right]} \label{eq:a5}
\end{equation}
 This function is sharply peaked around
$\kp=2 \pi \u/\rn$, and  only values of $\kp$ in the interval $\mid
\Delta \kp \mid < 1/(\rn \theta_{0 \nu})$ centered around this value
will contribute to the integral in equation (\ref{eq:a1}). It follows  
that the visibility $V(\u,\nu)$ picks up contributions  only from
those fluctuations $\Delta^{s}_{\HI}(\k,\nu)$ for which $\kp \approx 2
\pi \u/\rn$. 

A point to be noted is that in the Gaussian model for the beam pattern
its  Fourier transform is also  a Gaussian which is  a non-compact
function.  In reality, the Fourier transform of the beam pattern is
compact and it is zero outside a radius $D/\lambda$ {\it i.e.}
$a_{\nu}(\u-\frac{\kp \rn}{2 \pi})= 0$ for $\mid \u-\frac{\kp \rn}{2
  \pi} \mid > D/\lambda$. This fact plays a  role in the
later discussion and we shall refer back to it at the appropriate
place. 

In this paper we shall study in some detail the cross-correlation
$\langle V(\u_1,\nu_1) V^*(\u_2,\nu_2) \rangle$  between  the visibilities 
measured at two different baselines and  frequencies.  Using equations
(\ref{eq:a1}), (\ref{eq:a3}), (\ref{eq:a4}),   this turns out to be

\begin{eqnarray}
&& \langle V(\u_1,\nu_1) V^*(\u_2,\nu_2) \rangle = 
 \left[\bar{I} D b \right]_{1}  \left[\bar{I} D b \right]_2 
 \int \frac{d^3 k}{(2 \pi)^3}  P(k)
 \left[ 1 +  \beta_1 \frac{\kn^2}{k^2} \right]
\times  \label{eq:a6}  \, \, \\ &&
 \left[ 1 +  \beta_2 \frac{\kn^2}{k^2} \right]
\cos[\kn (r_1-r_2)]  \,  a_1(\u_1-\frac{\kp r_1}{2
  \pi}) a^*_2(\u_2-\frac{\kp r_2}{2 \pi}) \nonumber \,,
\end{eqnarray}
the imaginary part being zero. 
We shall  use equation (\ref{eq:a6}) to
calculate the correlations between the visibilities at different
baselines and frequencies. 
We next discuss  a few approximations which can be used to simplify
equation (\ref{eq:a6}) and which help in interpreting this equation.
We test the range of validity of these approximations and 
comment on this  later in the paper. 

Typically, the bandwidth $B$ is  small compared to the central
frequency $\nu_c$, and the quantities $I_{\nu}$, $b(z)$, $D(z)$,
$\beta(z)$, $\theta_{0\nu}$ and $\rn$ in equation (\ref{eq:a6}) do not
vary substantially across the band. We can evaluate these at $\nu_c$
instead of calculating them separately at $\nu_1$ and $\nu_2$, and at
some instances we 
do not explicitly show the subscript $\nu_1$ or $\nu_2$ for these
quantities. 
 The term $\cos[\kn (r_1-r_2)]$ is an exception  because $\kn
 (r_1-r_2)$ may be    large  even if $\Delta \nu =\nu2-\nu_1 \ll 
\nu_c$. This term may be approximated as  $\cos[\kn \rn^{'} \Delta
  \nu]$, where $\rn^{'}=d \rn/d \nu$. 

The term $  a_1(\u_1-\frac{\kp r}{2  \pi}) a^*_2(\u_2-\frac{\kp
  r}{2 \pi})$ in equation (\ref{eq:a6}) is a product of two
  functions, one sharply peaked at   $\kp = 2 \pi\u_1/r$  and
  another at $\kp=2 \pi \u_2/r$. The two 
  peaks have very little overlap if $\mid \u_1-\u_2\mid \gg
  1/\theta_{0}$, and the visibilities measured at  two such
  baselines   will be uncorrelated.  There will be  correlations only
  between   visibilities measured at different frequencies for nearby
  baselines, {\it ie.}   $\mid \u_1-\u_2\mid \leq 1/\theta_{0}$. For the
  situation where   $\u_1=\u_2 \gg   1/\theta_{0}$, 
  we may use the approximation (Paper II)
\begin{equation}  
\mid  a_1(\u_1-\frac{\kp r}{2  \pi})  \mid^2 
 \approx \left( \frac{2 \pi^3 \theta_{0}^2}{r^2} \right) \delta^2
\left( \kp- \frac{2 \pi}{r} \u_1 \right) 
\label{eq:a7}
\end{equation}
whereby  equation (\ref{eq:a6}) is considerably simplified and we have 
\begin{eqnarray}
\langle V(\u,\nu) V^{*}(\u,\nu+\Delta \nu) \rangle =
\frac{[\bar{I} b D \theta_{0}]^2}{2 r^2}
\int_0^{\infty} d \kn  P(k) \left[1+ \beta \frac{\kn^2}{k^2}
  \right]^2 \cos(\kn  r^{'} \Delta \nu) 
\label{eq:a8}
\end{eqnarray}
where $ k=\sqrt{(2 \, \pi \, U/r)^2  + \kn^2}$. 

Equation (\ref{eq:a8}) provides considerable insight on the
visibility-visibility cross-correlation in the situation where
$\u_1=\u_2=\u$. 
The point to note that the visibility signal measured at a baseline
$\u$ receives contributions only from the fluctuations $\Delta(\k)$
for which $\kp=(2 \, \pi \, \u)/r$, {\it ie.} $\kp$ is fixed but $\kn$
can have any value. As a consequence the correlation $\langle V(\u,\nu) 
V^*(\u,\nu+\Delta \nu)  \rangle$  directly probes the power spectrum 
$P(k)$ at all Fourier modes $k > k_{\rm min}\,   =2 \pi U/r$. 

In the next sections of this paper we use the formulas presented here
to investigate the HI signal expected for the GMRT and study how this
is 
related to the large-scale structures at the redshift where the HI
emission originated. We also compare the predictions of eq.
(\ref{eq:a6}) with those of (\ref{eq:a8}) and test the range of
validity of the approximations discussed above. 

\section{Some Useful  Parameters}
\subsection{GMRT}
The Giant Meterwave Radio Telescope (GMRT) has 30 radio antennas of
$45 {\rm m}$ diameter each. We present below, in tabular form,  some of
the parameters relevant  for the proposed HI 
observations.   We have  restricted our analysis to only two of the
GMRT frequencies which correspond to HI in the range $1 \le z \le
3.5$.  
\begin{center}
\begin{tabular}{|l|l|}
\hline
Frequencies \, $\nu_c$ (MHz) &  610 \, \,   325 \\
\hline
Possible Bandwidths - B (MHz) & 16 \,  8 \,  4 \,  2\,  1 \,  0.5 \,
\\ 
\hline
No of Channels - NC  & 128 \\
\hline
Antenna Separations - d (m) & $d_{\rm min} \sim 60 $ \,$ d_{\rm max}
\sim 25 \times 10^3$  \\
\hline
\end{tabular}
\end{center}
\begin{center}
\begin{tabular}{|c|c|c|c|}
\hline
\hspace{0.5cm} $\nu_c$ \hspace{0.6cm} & $\hspace{0.5cm} U_{\rm min}$
\hspace{0.5cm} &   \hspace{0.5cm} 
$U_{\rm max}$ \hspace{0.5cm} & \hspace{0.2cm} $\theta_0 \approx 
0.6 \times \theta_{\rm FWHM}$ \hspace{0.2cm}  \\
\hline
610 & 123 & $51 \times 10^3$ & $0.54^{\circ}$ \\
\hline
325 & 65 & $27 \times 10^3$ & $1.08^{\circ}$\\ 
\hline
\end{tabular}
\end{center}

\subsection{Cosmological Model}
For the background cosmological model we have used $H_0=100 \, h \,
{\rm km/s/Mpc}$ with $h=0.7$, $\Omega_{m0}=0.3$, 
$\Omega_{\Lambda0}=0.7$, $\Omega_{\rm Baryon\,0}=0.015 \, h^{-2}$  and 
$\Omega_{\rm   gas}=1. \times 10^{-3}$.   
The power spectrum of dark matter density fluctuations is normalized
to COBE (Bunn \& White 1996), and its shape is  determined using the  
analytic fitting form for the CDM power spectrum given by Efstathiou,
Bond and White (1992). The value of the shape parameter turns out to
be $\Gamma=0.2$ for the set of cosmological parameters used here.  
The power spectrum is shown in Figure~1. 

\begin{center}
\begin{figure}[h]
\mbox{\epsfig{file=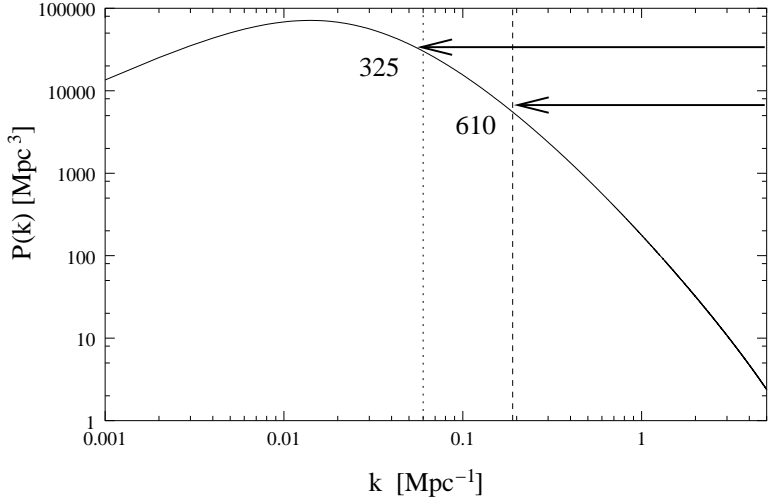,width=4.in,angle=0}}
\caption{This shows the part of the power spectrum  probed by the
  smallest baselines in GMRT for the two frequencies indicated  in the 
  figure. The power spectrum is normalised to COBE and is shown for
  $z=0$.}
\label{fig:1}
\end{figure}
\end{center}

Very little is known about the bias $b$ which relates the HI
fluctuations to the dark matter fluctuations, and we  have used a
fixed value $b=1$ throughout.  
\subsection{HI.}
We first  present the values of the redshift, comoving distance and the  
expected specific intensity of the HI distribution for the two GMRT
frequencies of our interest.  We also show values of $r^{'}
=dr/d\nu$ which are useful in calculating the comoving separation
corresponding to a given frequency interval {\it ie.} $\Delta r=r^{'} 
  \Delta \nu$. The last column in  the table below shows
the value of $k_{\rm min}$ for the  baseline $U=100$. 
This represents the smallest Fourier mode probed by this baseline. 
\begin{center}
\begin{tabular}{|c|c|c|c|c|c|}
\hline
$\nu_c$ & $z$ & $r$ (Mpc) & $\bar{I}$ (${\rm Jy/deg^2}$) & $ r^{'}$ (Mpc/MHz) &
$k_{\rm \min} \, ({\rm Mpc}^{-1})$\\  
\hline
610 & 1.33 & 4030 & $9.0 \times \, 10^{-4} $& 7.7 & 0.16 \\
\hline
325 & 3.37 & 6686 & $3.7 \times 10^{-4}$ & 11.3 & 0.09 \\
\hline
\end{tabular}
\end{center}

In Figure~1 we show the part of the power spectrum which will be probed by
$U_{\min}$ the smallest baseline available on GMRT. The value of
$k_{\rm min}$ will be higher for the larger baselines and these will
probe a smaller part of the power spectrum.  

\section{Results}
\subsection{Single Dish Observations}
It is possible to use the GMRT as a set of 30 single dish antennas
instead of an interferometric array. In such observations
the flux incident on each antenna is recorded and the signal from the
30 antennas is combined.  The flux $F(\nu)$ is  the visibility
$V(\u,\nu)$ at $\u=0$, and the flux expected from the redshifted HI
emission can be calculated  using  equation (\ref{eq:a1}). We use
equation  (\ref{eq:a6}) to calculate the  correlation between the flux
measured at different frequencies $\langle F(\nu) F(\nu + \Delta \nu)
\rangle$. The results for  the two GMRT bands
of interest.  are shown in figures 2 and 3.  The approximate formula
(eq. \ref{eq:a8}) is valid only when $\mid \u \mid \gg 1/\theta_0$,
and this cannot be used here. 
\begin{figure}[htb]
\begin{center}
\mbox{\epsfig{file=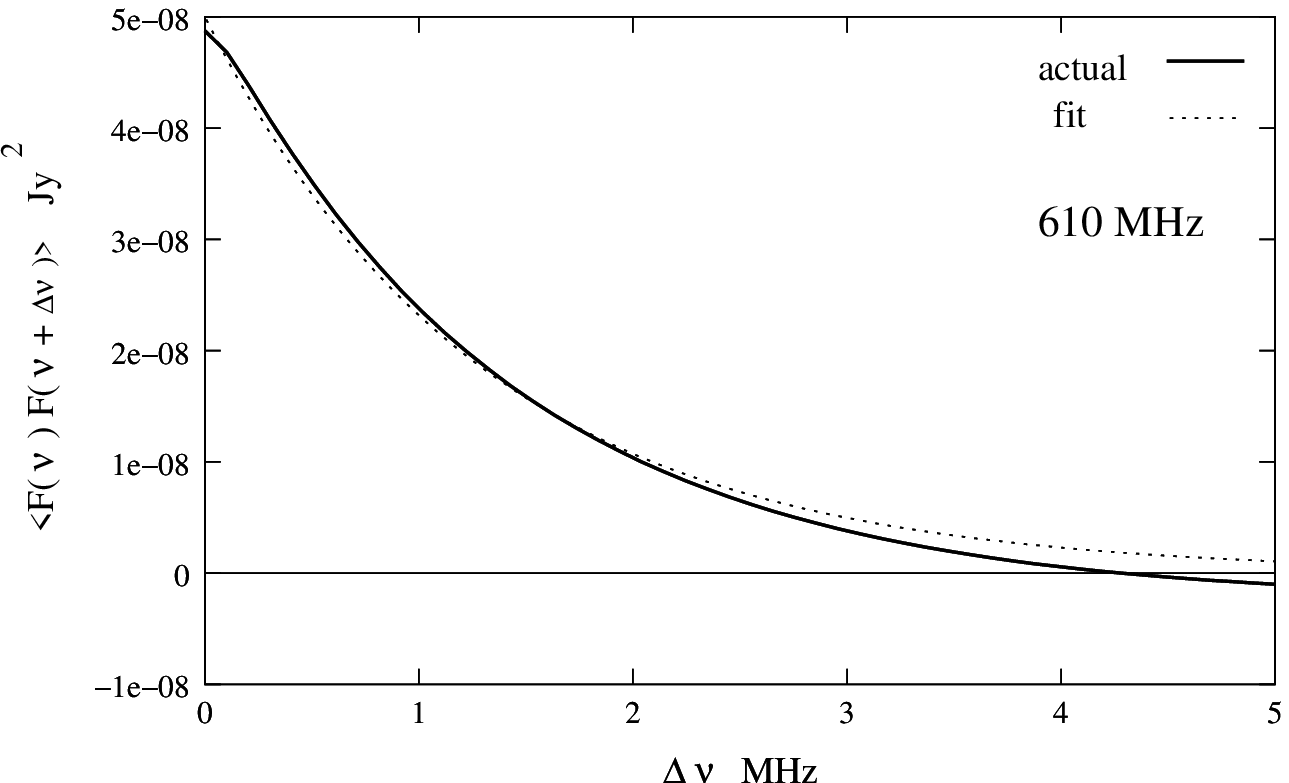,width=4.5in,angle=0}}
\end{center}
\caption{
This shows the  contribution from redshifted HI emission to
  the correlation between the flux expected at  two   different
  frequencies in the $610 {\rm   MHz}$ band for single
dish  observations using GMRT . The analytic  fit (eq.~\ref{eq:c1}) is
  also shown. } 
\label{fig:2}
\end{figure}
\begin{figure}[tbh]
\begin{center}
\mbox{\epsfig{file=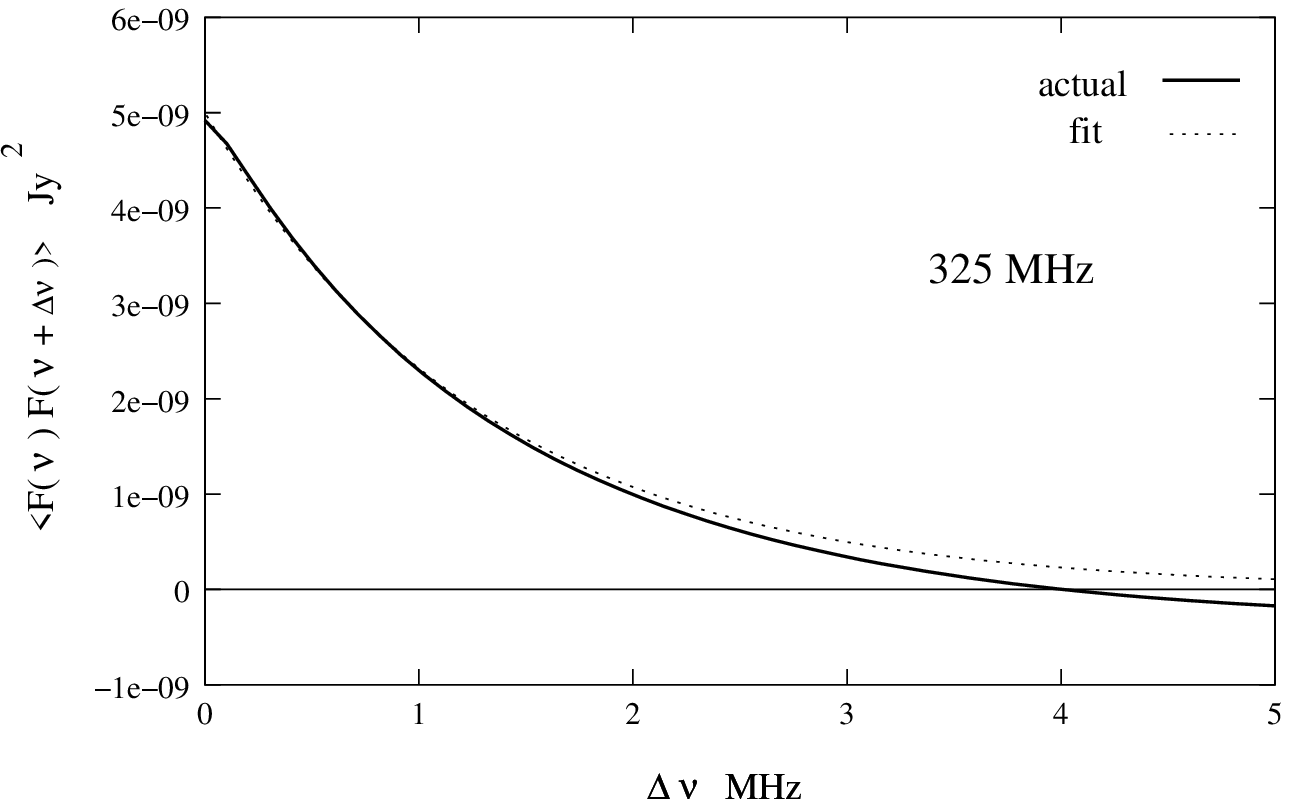,width=4.5in,angle=0}}
\end{center}
\caption{
This shows the  contribution from redshifted HI emission to
  the correlation between the flux expected at  two   different
  frequencies in the $325 {\rm   MHz}$ band for single
dish  observations using GMRT . The analytic  fit (eq.~\ref{eq:c1}) is
  also shown. } 
\label{fig:3}
\end{figure}

We find that the HI flux at neighbouring frequencies are
correlated. The correlation signal in the $610 {\rm MHz}$ band is
nearly an
order of magnitude higher than the correlation signal in the $325 {\rm
MHz}$ band.  In both the bands 
the correlation falls exponentially as  $\Delta \nu$ is
increased. For $\Delta \nu \le 2 {\rm MHz}$, the correlation can be
very well approximated by the fitting formula 
\begin{equation} 
\langle F(\nu) F(\nu + \Delta \nu) \rangle
=C_{\nu} \exp\left(\frac{-\Delta \nu}{1.3 {\rm MHz}} \right)
\hspace{1cm} \left\{
\begin{array}{c} 
C_{610}=5 \times 10^{-8} {\rm Jy}^2 \\ 
C_{325}=5 \times 10^{-9} {\rm Jy}^2 \\
\end{array}  \right. 
\label{eq:c1}
\end{equation}
The fitting formula (eq. \ref{eq:c1}) fails at larger $\Delta \nu$
where $\langle F(\nu) F(\nu + \Delta \nu) \rangle$ becomes
negative indicating that  the fluxes have a weak anti-correlation for
$\Delta \nu> 4 {\rm MHz}$. 

\subsection{Interferometric Observations}
We next consider interferometric observations with GMRT. The
correlation  $\langle V(\u_1,\nu_1) V^*(\u_2,\nu_2) \rangle$  is
maximum when $\u_1=\u_2$, and  we consider this situation first. 
In addition, it is a good approximation to represent the correlation as
a function of  $\Delta \nu=\nu_2-\nu_1$,  provided $\nu_1$ and $\nu_2$ are
in the same GMRT band, Our results are shown in figures 4. and 5.   
\begin{figure}[h]
\begin{center}
\mbox{\epsfig{file=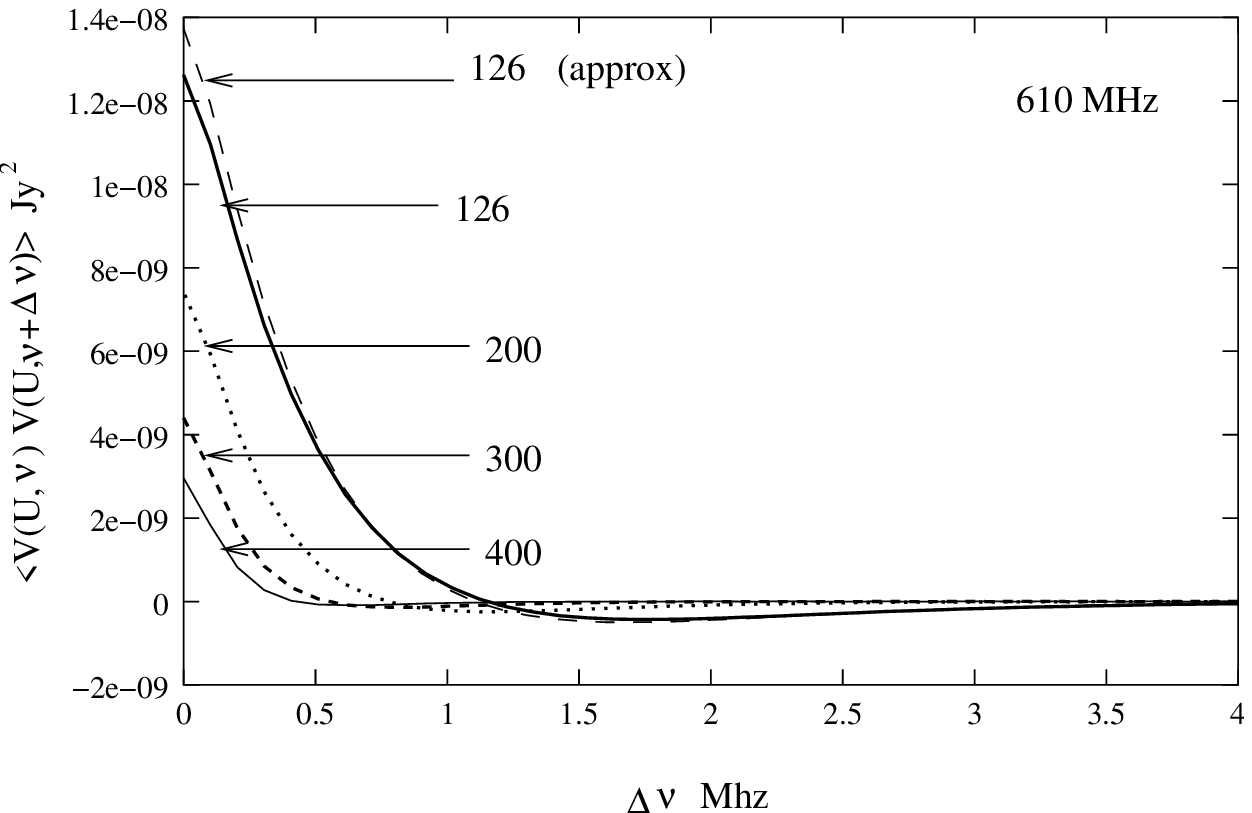,width=4.5in,angle=0}}
\end{center}
\caption{
This shows the correlation between the visibilities
$V(\u,\nu)$ and $V(\u,\nu+\Delta \nu)$  expected for the same baseline
  $\u$ at two different frequencies. The results are shown for
  different values of $U$ (shown in the figure)  starting from
  $U_{\rm min}$ for the $610 {\rm MHz}$ band. The results have been
  calculated using equation (\ref{eq:a6}), the predictions of the
  approximate formula (eq. (\ref{eq:a8}) are also shown for $U_{\rm min}$. 
}
\label{fig:4}
\end{figure}
\begin{figure}[h]
\begin{center}
\mbox{\epsfig{file=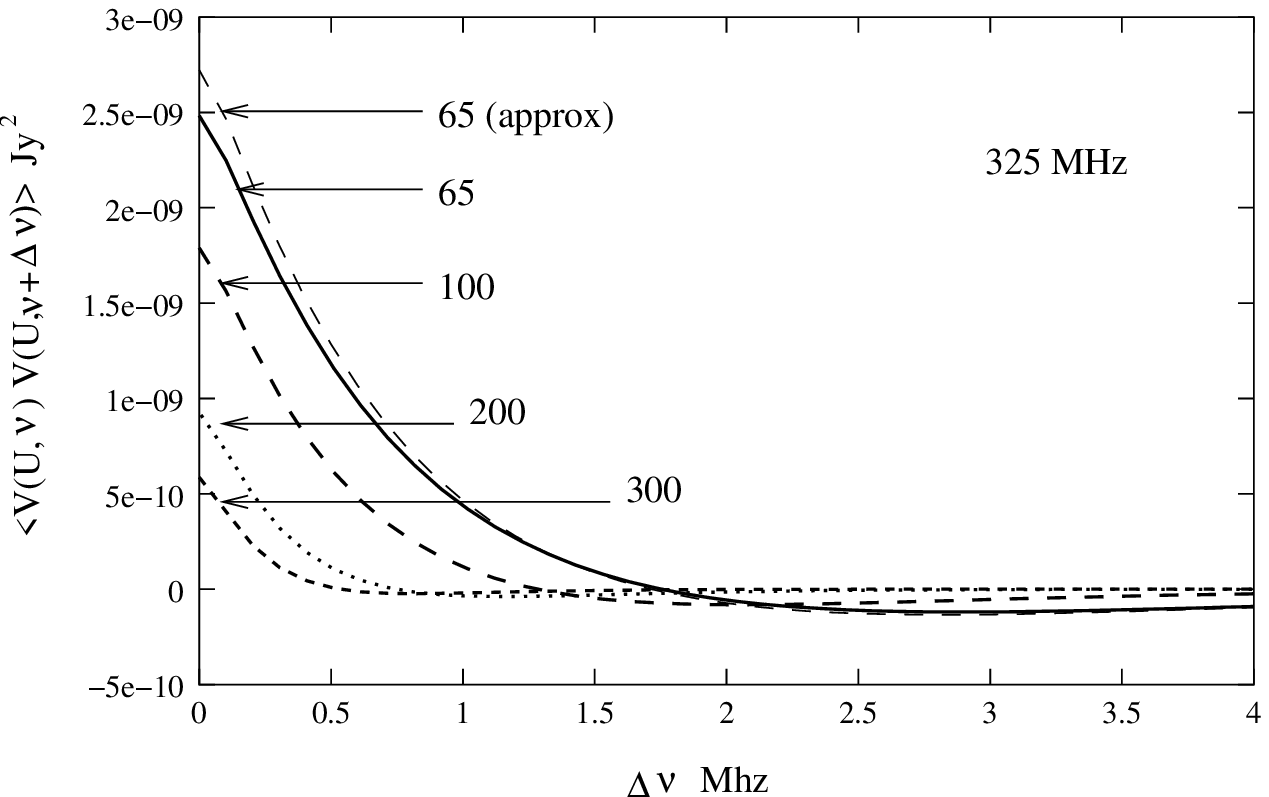,width=4.5in,angle=0}}
\end{center}
\caption{
This shows the same thing as figure 4. for the $325 {\rm
    MHz}$ band.}
\label{fig:5}
\end{figure}

We have calculated the correlations using two different formulas {\it
i.e.} equations (\ref{eq:a6}) and  equation (\ref{eq:a8}). The latter
 can only be used when $\u_1=\u_2=\u$,  and it is based on an
 approximation which is valid only when 
$\u \gg 1/\theta_0$.  We find that the approximate formula
(eq. \ref{eq:a8}) matches equation (\ref{eq:a6}) quite well ($\sim 10
\%$)  at the smallest baseline $U_{min}$ for both $610 {\rm MHz}$ and
$325 {\rm MHz}$ 
(figures 4 and 5). The agreement is better than $10 \%$ for the larger
baselines.  

The correlation signal is the strongest at the smallest
baseline $U_{\min}$ and falls  quickly for the larger values of
$U$. The larger baselines receive contributions from a smaller part of
the power spectrum (figure 1.) and the fall-off with $U$ reflects the
shape of the power spectrum at the mode $k \sim (2 \pi U/r)$. At a
fixed $U$, the visibilities at adjacent frequencies are correlated,
the correlation falls  approximately exponentially with increasing
$\Delta \nu$. At the larger  baselines the correlation decreases faster
with increasing $\Delta \nu$. At all baselines, the correlation crosses
zero and there are  anti-correlations for large $\Delta \nu$. We find
that in the range  of $U$ and $\Delta \nu$  where the
visibility-visibility correlation signal is large, it  can
be approximated by the fitting formula  
\begin{eqnarray}
\langle V(U,\nu) V^{*}(U,\nu+\Delta \nu) \rangle = 
C_{\nu}
\exp\left( \frac{- \Delta \nu \, u^{0.8}}{0.7}  \right)
\frac{\sin\left(2 \, \Delta \nu \, u^{1.2} \right)}{2\,  \Delta \nu \,
  u^{1.2}}  \\ \nonumber \\
C_{610}=1.6 \times 10^{-8} {\rm Jy}^2 u^{-1.2} \hspace{2cm}
C_{325}=1.8 \times 10^{-9} {\rm Jy}^2 u^{-0.9} \nonumber 
\end{eqnarray}
where 
where $u=(U/100)$ and $\Delta \nu$ is  in MHz.

We next calculate the correlation 
$\langle V(\u,\nu) V^*(\u+\Delta \u ,\nu + \Delta \nu ) \rangle$
between the visibilities at two different baselines and
frequencies. We use equation (\ref{eq:a6}) to calculate this.  In
general, the behaviour of the correlation will depend on both,  the
magnitude and the  direction of $\Delta \u$. We have separately
considered two situations (1.) $\Delta \u$ parallel to $\u$ and (2.) 
 $\Delta \u$ perpendicular  to $\u$.  The results  are presented in
figures 6 and 7 for $U=200$. The results are similar for   other
values of $U$.    
\begin{figure}[h]
\begin{center}
\mbox{\epsfig{file=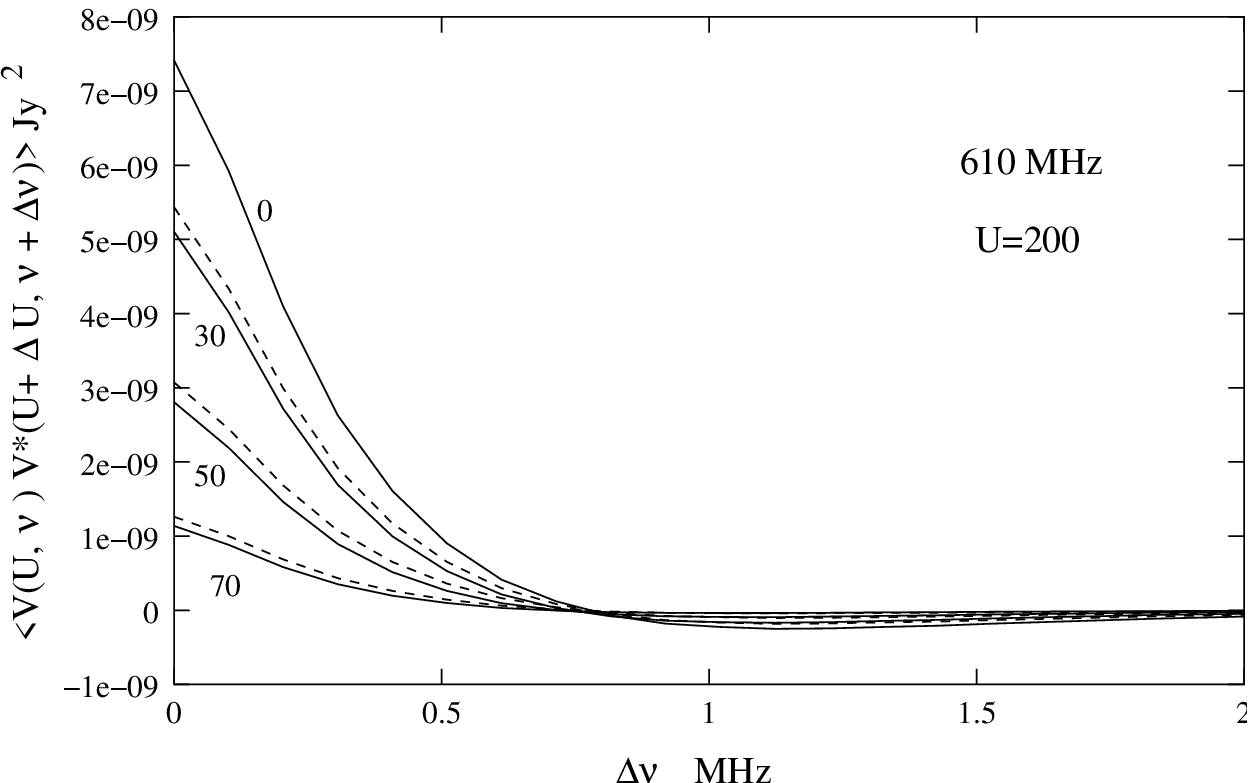,width=4.5in,angle=0}}
\end{center}
\caption{
This shows the expected correlation between the visibilities
$V(\u,\nu)$ and $V(\u + \Delta \u,\nu+\Delta \nu)$  at two different
  baselines and  frequencies. The results are shown at
$U=200$ for different values of $\Delta U$ (shown in the figure), when
  $\Delta \u$ is parallel to $\u$ (solid line) and   $\Delta \u$ is
  perpendicular to $\u$ (dashed line). These results are for the $610
  {\rm MHz}$ band.}  
\label{fig:6}
\end{figure}
\begin{figure}[h]
\begin{center}
\mbox{\epsfig{file=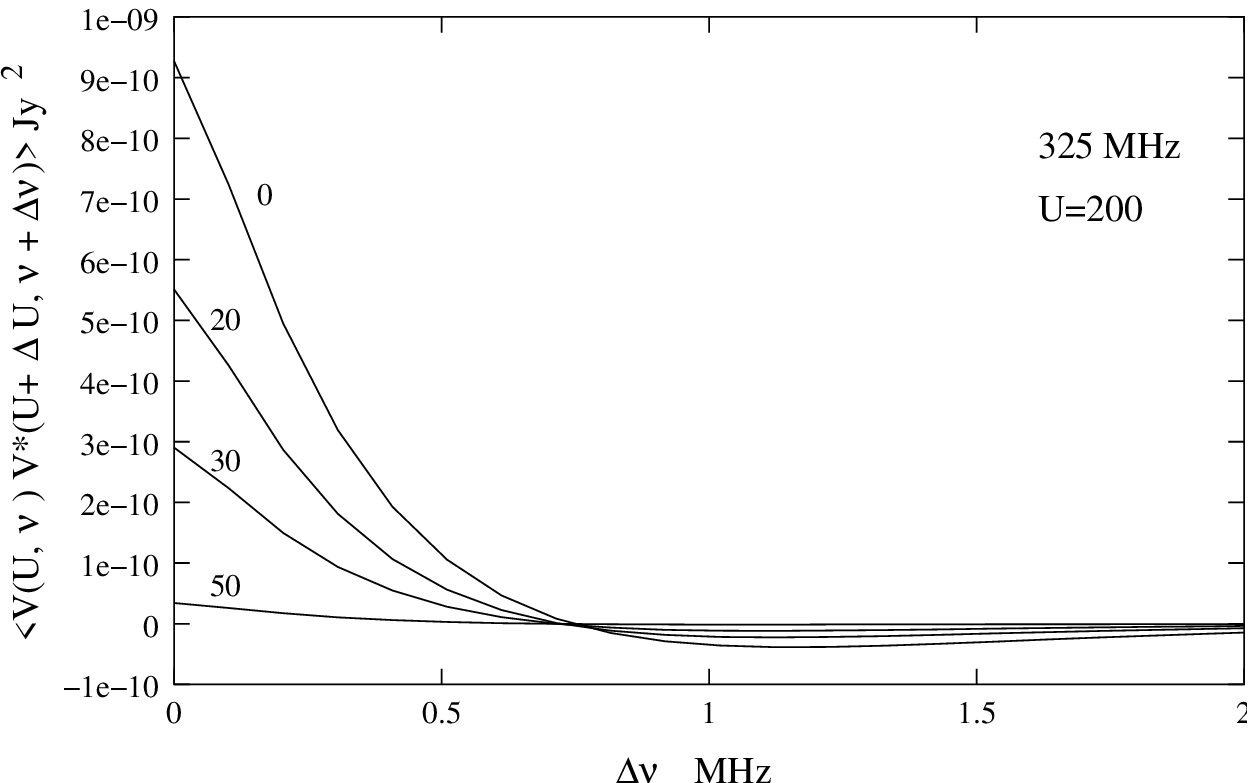,width=4.5in,angle=0}}
\end{center}
\caption{
This is same as figure 6 except that it is for the  $325
  {\rm MHz}$ band. Here the results do not depend on whether  $\Delta \u$ is
parallel or perpendicular to $\u$. }
\label{fig:7}
\end{figure}

We find that the  correlation falls rapidly as $\Delta \u$
increases. The  $\Delta \u$ dependence is found to be  anisotropic for
the $610 \, {\rm MHz}$ band, . The correlation falls faster when
$\Delta \u$ is parallel to $\u$ than when it is perpendicular to
$\u$. The $\Delta \u$ dependence is found to be isotropic for the $325
\, {\rm MHz}$  band,  In both the bands the  $\Delta \u$ dependence
can be well approximated by a  Gaussian  
\begin{eqnarray}
\langle V(\u,\nu) V^*(\u+\Delta \u ,\nu + \Delta \nu ) \rangle &=&
\exp\left[ -
  \left(\frac{\Delta U_{\parallel}}{a_{\nu}} \right)^2 - 
  \left(\frac{\Delta U_{\perp}}{b_{\nu}} \right)^2 \right] 
\times \nonumber \\
&& \langle
V(U,\nu) V^*(U,\nu + \Delta \nu ) \rangle \label{eq:b11} \\
 \, a_{610}=50, \,  b_{610}=53 \hspace{2cm} &&
 a_{325}= b_{325}=28 \nonumber 
\end{eqnarray}
where $\Delta U_{\parallel}$ is the component of $\Delta \u$ parallel
to $\u$, and $\Delta U_{\perp}$ is the component perpendicular to
$\u$. This fitting function is found to be a very good approximation
over a large range of of $\u$, $\Delta \u$ and $\Delta \nu$.

It should be noted that contrary to the predictions of our fitting
formula (eq. \ref{eq:b11}),  the  cross-correlations between
the different baselines is {\it exactly zero} {\it i.e.} $\langle
V(\u,\nu) V^*(\u+\Delta \u  ,\nu + \Delta \nu ) \rangle =0$ if $\mid
\Delta \u \mid > D/\lambda$.  This is a consequence of the fact that
we have modeled the  Fourier transform of the beam pattern as a
Gaussian whereas it actually has compact support (discussed in
Section2.), The cross-correlations are zero for $\mid  \Delta \u
\mid \, > \, 90$ at $610\,  {\rm MHz}$  and $\mid \Delta \u \mid \, >
\, 50$ at $325 \, {\rm MHz}$. This discrepancy does not significantly 
change our results as the Gaussin function in equation (\ref{eq:b11})
is extremely  small ($< 1 \%$) at the values of $\mid \Delta \u
\mid$ where the cross-correlations vanish. 
\section{Discussion and Conclusions.}
We have investigated in detail the redshifted HI emission signal
expected in GMRT observations in the $325 {\rm MHz}$ and the $610 {\rm
  MHz}$ frequency bands. The properties of the
visibility-visibility cross-correlation  $\langle V(\u,\nu)
V^*(\u+\Delta \u,\nu + \Delta \nu)\rangle$ which  we propose
as a statistical estimator for detecting and analyzing the
signal have been studied.  The signal is maximum at $U_{\rm min}$, and
we find that  
$\langle V(\u_{\rm min},\nu) V^{*}(\u_{\rm min},\nu)\rangle$ is $\sim
1.3 \times 10^{-8} {\rm Jy}^2$ at $610 \,  {\rm MHz}$ and $\sim 2.5 \times
10^{-9} {\rm Jy}^2$ at $325 \,  {\rm MHz}$. The signal falls at larger
baselines, and it drops by an order of magnitude by $U \sim 10 \times
U_{\rm min}$. Fourteen of the thirty GMRT antennas are located within
$1 {\rm km}$. These baselines could be used for detecting the
signal whereas the larger baselines where there is very little signal 
could be used as a control to test that what we detect is actually the
HI signal and not an artifact. 

The correlation   $\langle V(\u,\nu) V^*(\u,\nu + \Delta \nu)\rangle$ 
fall exponentially with increasing 
$\Delta \nu$, the width being around $0.5 \, {\rm MHz}$ at $U_{\rm
  min}$.  The correlations fall faster at the larger baselines. The
$\Delta \nu$ dependence of the correlation  will be crucial in
detecting the HI signal, for neither the  system noise nor the
galactic/extragalactic  continuum radiation are  expected to have such
a feature.  

The correlation  $\langle V(\u,\nu) V^*(\u+\Delta \u,\nu + \Delta
\nu)\rangle$ falls rapidly with increasing $\Delta \u$, the $\Delta
\u$ dependence being well described by a  Gaussian. The correlation
falls by $50 \%$ when 
$\mid \Delta \u \mid \sim 42$ at $610 {\rm MHz}$ and  $\mid \Delta \u
\mid \sim 23$ at $325 {\rm MHz}$. The correlation falls by $10 \%$
when $\mid \Delta \u \mid \sim 16$ at $610 {\rm MHz}$ and  $\mid \Delta \u
\mid \sim 9$ at $325 {\rm MHz}$.
The $\Delta \u$ dependence plays an
important role in three different contexts in this discussion. 
\begin{itemize}
\item[a.]  For a pair of antennas at a separation $\d$, the
  value  $\u=\d/\lambda$    changes as   $\nu$ varies across  the
  frequency band.   The correlation between the visibilities measured
  by   this pair of antennas at two different frequencies will be  
$\langle V(\u,\nu) V^*(\u+ \Delta \u,\nu + \Delta \nu)\rangle$, where
  $\Delta \u=-(\Delta \nu/\nu) \u$ arises due to the change in
  wavelength. . For  $U \le 500 $ and $\Delta
  \nu \le 1 {\rm MHz}$, which is the range where there is significant 
  correlation,   we find  $\mid \Delta \u \mid < 2$. Our
  investigations show   that this 
  effect will cause the correlation to fall by a factor less than $10
  \%$ from   $\langle V(\u,\nu) V^*(\u,\nu + \Delta \nu)\rangle$, and
  we can ignore the change in $\u$ due to the change in frequency. 
\item[b.] The rotation of the earth causes all the baselines to change
  in time and for a time interval $\Delta t$ we have $\Delta
  \u=\Delta t \, \om {\bf  \times} \u$. This has to be taken into
  account when deciding on the time interval $\Delta t$ over which
  the visibility can be integrated without its value changing
  significantly. Our analysis shows that the correlation in the
  visibility falls by $10 \%$ or less if $\Delta U   \le 10$ and we
  use this as a  the criteria  to determine $\Delta t$. In the range
  $U \le   500$ where we expect most of the signal we obtain $\Delta t 
  \sim 5\,   {\rm min}$. 
\item[c.] As the earth rotates, GMRT observations will produce
  visibility data for a very  large number of baselines $\u$. The
  values of $\u$ will not be   uniformly distributed in the plane of
  the array,  and there will be regions which are more densely sampled
  than others. It will be useful to reduce the volume of data by
  combining  the data in the densely   sampled regions and thereby
  produce a set of values of the visibility on a 
  regular grid of $\u$ values.   Our analysis shows that the grid has
  to be at an interval$\Delta U$  $\sim 40$ at $610 {\rm MHz}$ and
  $\sim 25$ at $325  {\rm MHz}$, the visibility signal gets
  uncorrelated after this interval. 
\end{itemize}

Two issues, the system noise and the contribution of the galactic and
extragalactic continuum sources have not been addressed here. Further,
we have modeled the HI gas as having a continuous distribution,
whereas in reality the HI gas is in discrete, small
clouds. Investigations are currently underway on these issues and the
results will be communicated in forthcoming publications. 

Finally, the visibility-visibility correlation signal depends
the equation of state of the dark matter and the dark energy at the
the epoch when the HI emission originated. This dependence comes in
through the redshift space distortion parameter, the comoving distance
and its derivative,  all of which depend on the equation of state of
the universe. We shall address  the possibility of using HI
observations to probe the equation of state of the universe 
at at high redshifts in a future paper. 

{\it Acknowledgment.} SB would like to than Jayaram N Chengalur and 
Shiv K Sethi for useful discussions. SB would also like to acknowledge
financial support from BRNS, DAE, Govt. of India, for financial
support through sanction No. 2002/37/25/BRNS. SKP would like to
acknowledge the Associate Program, IUCAA for   supporting his visit to
IIT,Kgp and CTS, IIT Kgp for the use of its facilities.  \\
\vspace{0.5cm}

\underline{References}

\begin{itemize}
\item[] Bharadwaj S., Nath B. \& Sethi S.K. 2001, JAA. 22, 21
\item[] Bharadwaj, S.~\&  Sethi, S.~K.\ 2001, JAA, 22, 293  
\item[] Bunn E. F. \&  White M. 1996,  ApJ, 460, 1071
\item[] Efstathiou, G., Bond, J. R.  \& White, S. D. M. 1992,
 MNRAS, 250, 1p
\item[] Lahav O., Lilje P. B., Primack J. R. and Rees M., 1991, MNRAS, 251, 128
\item[] Lanzetta, K. M., Wolfe, A. M., Turnshek, D. A. 1995,
    ApJ, 430, 435
\item[] Peebles, P. J. E. 1980, {\it The Large-Scale Structure of
    the Universe \/}, Princeton, Princeton University Press
\item[] P\'eroux, C., McMahon, R. G., Storrie-Lombardi,
    L. J. \& Irwin, M .J. 2001, astro-ph/0107045
\item[]  Saini, T., Bharadwaj, S. \& Sethi, K. S. 2001, ApJ, 557, 421
\item[] Storrie--Lombardi, L.J., McMahon, R.G., Irwin, M.J. 1996, MNRAS,
   283, L79
\item[] Swarup, G., Ananthakrishan, S., Kapahi, V. K., Rao, A. P.,
 Subrahmanya, C. R., \& Kulkarni, V. K. 1990, Curr. Sci., 60, 95
\end{itemize}

\end{document}